\documentclass[11pt]{article}
\usepackage[letterpaper,hmargin=1in,vmargin=1in]{geometry}
\usepackage{amsfonts}

\newtheorem{definition}{Definition}[section]
\newtheorem{theorem}{Theorem}[section]

\newtheorem{corollary}[theorem]{Corollary}
\newenvironment{proof}{\par \noindent
            {\bf Proof. \hspace{1mm}}}{\hfill$\Box$ \vspace*{3mm}}

\begin{document}
\begin{center}
{\Large\bf Security Notions for Quantum Public-Key Cryptography}\\[2em]
{\large Takeshi Koshiba}\\[1em]

Area of Informatics, Division of Mathematics, Electronics, and Informatics,\\
Graduate School of Science and Engineering, Saitama University\\
255 Shimo-Okubo, Sakura-ku, Saitama 338-8570, Japan.\\
{\tt Email: koshiba@acm.org}
\end{center}
\vskip 2em
\begin{quote}
{\bf Abstract: }
It is well known that Shor's quantum algorithm for integer factorization
can break down the RSA public-key cryptosystem, which is widely used
in many cryptographic applications. 
Thus, public-key cryptosystems in the quantum computational setting
are longed for cryptology.
In order to define the security notions
of public-key cryptosystems, we have to model the power of the sender, 
receiver, adversary and channel. 
While we may consider a setting where quantum computers are
available only to adversaries,
we generally discuss what are the right security notions 
for (quantum) public-key cryptosystems in the quantum computational setting.
Moreover, we consider the security of quantum public-key cryptosystems
known so far.\\[1em]
\end{quote}

\section{Introduction}
Shor's quantum algorithm \cite{Shor97} for the integer factorization problem can
break down the RSA cryptosystem. This fact may be seen 
as a negative aspect of the power of the quantum mechanism.
On the other hand, a quantum key distribution protocol due to 
Bennett and Brassard \cite{BB84} is one of the most successful cryptographic systems.
It is natural to consider that we can defend even public-key cryptosystems
by using quantum computers against the quantum adversary.
Since lattice-based cryptosystems such as 
the Ajtai-Dwork public-key cryptosystem \cite{AD97}
are based on the computational hardness of the shortest vector problem (SVP)
that is not known to be efficiently solvable by using quantum computers,
they have attracted researchers' attention (\cite{GGH97,Regev04a,Regev04b,Regev05})D

Basically, the lattice-based cryptosystems are classical ones that are
likely to withstand quantum adversaries. On the other hands, there are
public-key cryptosystems in which the power of quantum computation is
ingeniously applied. The first quantum public-key cryptosystem
was proposed by Okamoto, Tanaka and Uchiyama \cite{OTU00}.
The Okamoto-Tanaka-Uchiyama system (OTU00 system, for short)
is one of knapsack-based cryptosystems, which
are based on the hardness of some subproblems of the NP-complete knapsack
problem (or the subset sum problem). Generally speaking, the security
of cryptosystems is related to the average-case complexity of their underlying
problems and some of knapsack-based cryptosystems have been actually broken.
Early knapsack-based cryptosystems utilized hidden linear relations between
the public-key and the secret-key and the attack algorithms could efficiently
find the hidden linear relations. Chor and Rivest \cite{CR88}
incorporated ``easily solvable'' discrete logarithmic relations 
into the key generation in order to prevent the above attack algorithms
and proposed a knapsack-based cryptosystem. OTU00 system can be seen as
an extension of the Chor-Rivest system since ``arbitrary'' discrete logarithmic
relations can be introduced in the key generation of OTU00 system 
by using Shor's algorithm.

The OTU00 system requires that receivers have quantum computers
to generate keys and the other information is totally classical.
On the other hand, Kawachi, Koshiba, Nishimura and Yamakami \cite{KKNY05} 
proposed a quantum public-key cryptosystem where the parties concerned
including the adversary and the channel are quantum. 
The Kawachi-Koshiba-Nishimura-Yamakami system (KKNY05 system, for short)
is the first provably-secure quantum public-key cryptosystem
(of the indistinguishability property).
In \cite{KKNY05}, it is shown that if there exists an efficient quantum 
algorithm to break the KKNY05 system then the graph automorphism 
can be efficiently solvable even in the worst-case. The KKNY05 system
can be seen as an quantum extension of Goldwasser-Micali's probabilistic
encryption system \cite{gm84}. In \cite{gm84}, they introduced two
security notions for public-key cryptosystems: the indistinguishability
(a.k.a. polynomial security) and the semantic security. Their
probabilistic encryption was shown to have the indistinguishability in \cite{gm84}
and later shown to be semantically secure as a consequence of the equivalence
between the indistinguishability and the semantic security
\cite{micali1}.

In this paper, we discuss the appropriate definitions of security
notions for quantum public-key cryptosystems and derive relations among them.
In the case of classical public-key cryptosystems, security notions
are defined in terms of the adversary model and the goal of the adversary \cite{bellare19}.
As adversary models, there are ciphertext only attack,
chosen plaintext attack and (non-adaptive/adaptive) chosen ciphertext attack.
As goals of the adversary, we usually consider the one-wayness (of encryption
function), the indistinguishability (of encrypted messages),
semantic security \cite{gm84} and non-malleability \cite{dolev2}.
For example, the ElGamal public-key cryptosystem \cite{elgamal}
is one-way (against chosen plaintext attack) on the assumption that 
the computational Diffie-Hellman problem \cite{diffie}
is hard, and semantically secure (against chosen plaintext attack)
on the assumption that the decisional Diffie-Hellman problem is hard.
Moreover, the Cramer-Shoup public-key cryptosystem \cite{CS03},
an extension of the ElGamal cryptosystem, is shown to be non-malleable
against adaptive chosen ciphertext attack on the assumption that 
the decisional Diffie-Hellman problem is hard and there exists a family
of universal one-way hash functions.

As mentioned, we discuss the validity of the analogous definitions
of security notions for quantum public-key cryptosystems. The discussion
involves the compatibility with the non-cloning theorem and the difference
between quantum and classical leakage of information from ciphertext.
(For example, it is known that the imperfect randomness alters
security notions of public-key cryptography \cite{koshiba01,koshiba02}.)
In this paper, we consider how to define security notions
for quantum public-key cryptography and especially revisited
the definition of the event such that no information is leaked from ciphertext.
Specifically speaking, we give an analogous definition of
the indistinguishability for quantum public-key cryptosystems and
two definitions of the semantic security both from viewpoints of
the classical and quantum leakage and show the equivalence 
(against chosen plaintext attack) among them.
We also give a quantum definition of non-malleability and show
the equivalence between the indistinguishability and the non-malleability
against chosen ciphertext attack. As a corollary of the equivalence,
we show that the KKNY05 system is semantically secure.

\section{Security Notions of Classical Public-Key Cryptography}
Before considering quantum public-key cryptosystems, we review classical
public-key cryptosystems and their security notions.
\begin{definition}\rm
A {\em public-key cryptosystem\/} is described by a quadruple $(G,M,E,D)$.
Each component is defined as follows.
\begin{enumerate}\itemsep=0pt\parsep=0pt
\item A {\em key generation algorithm\/} $G$ is a probabilistic polynomial-time
algorithm that on input $1^n$ (the unary representation of $n$) outputs
a pair $(e,d)$ of strings, where $e$ is called {\em encryption key\/}
and $d$ {\em decryption key}.
\item $M=\{M_n\}_{n\ge 1}$ is a family of plaintext spaces to which every plaintext
message belongs. We also assume that the description of $M_n$ can be
output by a polynomial-time (uniform) algorithm.
\item For every $n$ and for every $(e,d)\in {\rm supp}(G(1^n))$ and 
$\alpha\in M_n$, a probabilistic polynomial-time {\em encryption algorithm\/}
$E$ and a deterministic polynomial-time {\em decryption algorithm\/} $D$ satisfy
that $\Pr\limits_{E}[D(d,E(e,\alpha))=\alpha]=1$.
\end{enumerate}
The integer $n$ is sometimes called {\em security parameter}.
The string $E(e,\alpha)$ is {\em encryption\/} of a plaintext $\alpha\in M_n$ 
with respect to the encryption key $e$, and
$D(d,\beta)$ is {\em decryption\/} of a ciphertext $\beta$ with respect to
the decryption key $d$. Though encryption and decryption keys are fed to
the encryption algorithm $E$ and the decryption algorithm $D$ respectively,
we sometime use the notation $E_e(\cdot)$ and $D_d(\cdot)$.
\end{definition}

Before reviewing security notions for public-key cryptosystems,
we consider how ingredients to define the security notions should be.

\subsection*{Attack Models}
Attack models on public-key cryptosystems in general consist of two phases.
In the first phase, an adversary is given the public-key $e$ where
$(e,d)$ is distributed according to $G(1^n)$.
In the second phase, the adversary is given a challenge ciphertext.
As attack models, ciphertext only attack (COA), chosen plaintext attack (CPA) and
chosen ciphertext attack (CCA) are well known. The distinct attribute among them
is what oracle the adversary can invoke. The adversary in the COA model cannot
invoke any oracles. The adversary in the CPA model can access to the oracle
that can reply to the query of plaintext with the corresponding ciphertext.
In the classical setting, the COA and CPA models are equivalent because
the adversary can encrypt any chosen message by himself using the public-key.
Since the relation between the COA and CPA models in the quantum setting is subtle, 
we postpone this issue to the next section.
The adversary in the CCA model can access to the oracle
that can reply to the query of ciphertext with the corresponding plaintext.
If the adversary can invoke the CCA oracle only in the first phase, then
the model is called non-adaptive CCA or CCA1. If the adversary can invoke
the CCA oracle in the both phases, the model is called adaptive CCA or CCA2.
Note that there is a limitation on the oracle invocation in the second phase,
where the adversary is not allowed to ask the challenge ciphertext. Otherwise,
it does not make sense. 

\subsection*{Computational Models of Adversaries}
In general, the adversary is modeled by a probabilistic polynomial-time
Turing machine as legitimate users are. We sometimes adopt polynomial-size
circuit family as a bit stronger computational model of the adversary.
In this paper, we define the security notions in terms of the non-uniform
computational model. Though technical difficulties between uniform models
and non-uniform models may be quite different, we adopt the non-uniform
model to simply the discussion. That enables us to grasp the 
security notions essentially.
More details on the uniform and the non-uniform models in cryptography
can be found in \cite{goldreich04}.

\subsection*{Goals of Adversaries}
Security notions are defined by determining when we say that
the adversary succeeds in the attack. As goals of the adversary,
the one-wayness (of the encryption), the indistinguishability
(of ciphertext), semantic security and non-malleability have been
well studied.

Before giving the definitions, we prepare some terminology.
A function $\mu: {\mathbb N}\rightarrow {\mathbb R}$ is {\em negligible with 
respect to\/} $k$ if $\mu(k)<1/p(k)$ for every positive polynomial $p(\cdot)$ 
and for every sufficiently large $k$. 
A function $\nu:\{0,1\}^\ast\rightarrow \{0,1\}^\ast$ is {\em polynomially
bounded\/} if there exists a polynomial $q(\cdot)$ such that
$|\nu(x)|\le q(k)$ for every $k$ and $x\in \{0,1\}^k$.

\begin{definition}\rm
A (classical) public-key cryptosystem $(G,M,E,D)$ is {\em one-way} if
for every family $\{C_n\}_{n\ge 1}$ of polynomial-size circuits,
\[ %\begin{array}{l}
\Pr\limits_{G,E,X_n}\Bigl[
C_n(e,E_e(\alpha))=\alpha\,\big|
(e,d)\leftarrow G(1^n);\alpha\leftarrow X_n
\Bigr] - \frac{1}{|M_n|}
 \]
is negligible with respect to $n$, where $X_n$ is the uniform distribution on $M_n$.
\end{definition}
{\bf Remark.} The above definition is slightly different from the standard
one. This is because the size of $M_n$ does not always depend on $n$.
Typically, we may consider the case $M_n=\{0,1\}$.
Moreover, $C_n$ can invoke oracles corresponding to the attack model.
Otherwise stated, we do not explicitly describe the oracle.

\begin{definition}\rm
A public-key cryptosystem $(G,M,E,D)$ has the {\em indistinguishability\/}
if for every family $\{C_n\}_{n\ge 1}$ of polynomial-size circuits,
every positive polynomial $p$, sufficiently large $n$, every $x,y\in M_n$,
\[ %\begin{array}{l}\displaystyle
\Bigr| \Pr\limits_{G,E}\bigl[C_n(e,E_e(x))=1\mid (e,d)\leftarrow G(1^n)
\bigr] -
\Pr\limits_{G,E}
\bigl[C_n(e,E_e(y))=1\mid (e,d)\leftarrow G(1^n)\bigr] \Bigl| < \frac{1}{p(n)}.
\]
\end{definition}

\begin{definition}\rm
A public-key cryptosystem $(G,M,E,D)$ is {\em semantically secure\/} if
there exists a (uniform) probabilistic polynomial-time computable 
transformation $T$ such that for every family $\{C_n\}_{n\ge 1}$
of polynomial-size circuits, 
every probability ensemble $\{X_n\}_{n\ge 1}$ each on $M_n$, 
every pair of polynomially bounded functions $f,h:\{0,1\}^*\rightarrow \{0,1\}^*$,
\[ \begin{array}{l}
\Pr\limits_{G,E,X_n}\Bigl[
C_n(e,E_e(\alpha),h(\alpha))=f(\alpha)\,\big|
(e,d)\leftarrow G(1^n);\alpha\leftarrow X_n
\Bigr]\\\displaystyle
\hspace*{5mm}- \Pr\limits_{T,G,X_n}
\Bigl[C_n'(e,h(\alpha))=f(\alpha)\,\big|
(e,d)\leftarrow G(1^n);\alpha\leftarrow X_n
\Bigr] 
\end{array} \]
is negligible with respect to $n$, where $C_n'=T(C_n)$.
\end{definition}
{\bf Remark.} Functions $f,h$ in the above are not necessarily recursive.

\begin{definition}\rm
A public-key cryptosystem $(G,M,E,D)$ is {\em non-malleable\/} if
there exists a (uniform) probabilistic polynomial-time computable 
transformation $T$ such that for every family $\{C_n\}_{n\ge 1}$
of polynomial-size circuits, 
every probability ensemble $\{X_n\}_{n\ge 1}$ each on $M_n$, 
every polynomially bounded function $h:\{0,1\}^*\rightarrow \{0,1\}^*$,
every relation $R$ computable by a family of polynomial-size circuits,
\[ \begin{array}{l}
\Pr\limits_{G,E,X_n}\Bigl[
C_n(e,E_e(\alpha),h(X_n))=E_e(\alpha')
\land (\alpha,\alpha')\in R\,\big|\,
(e,d)\leftarrow G(1^n),\alpha\leftarrow X_n
\Bigr]\\\displaystyle
\hspace*{5mm}- \Pr\limits_{T,G,X_n}
\Bigl[C_n'(e,h(\alpha))=E_e(\alpha')
\land (\alpha,\alpha')\in R\,\big|\,
(e,d)\leftarrow G(1^n),\alpha\leftarrow X_n
\Bigr] 
\end{array} \]
is negligible with respect to $n$, where $C_n'=T(C_n)$.
\end{definition}

For these security notions, the following theorems hold.

\begin{theorem}[\cite{gm84,micali1}]
A public-key cryptosystem $(G,M,E,D)$ is semantically secure
against the chosen plaintext attack if and only if $(G,M,E,D)$
has the indistinguishability against the chosen plaintext attack.
\end{theorem}

\begin{theorem}[\cite{dolev2}]
A public-key cryptosystem $(G,M,E,D)$ is non-malleable
against the adaptive chosen ciphertext attack if and only if $(G,M,E,D)$
has the indistinguishability against the chosen plaintext attack.
\end{theorem}

Other relations can be found in \cite{bellare19} though definitions
are slightly different from ours.

\section{Security Notions for Quantum Public-Key Cryptography}
We begin with a definition of quantum public-key cryptosystem, which
is a generalization of classical public-key cryptosystem. In this paper,
we focus on cryptosystems in which all plaintext messages are classical.

\begin{definition}\rm
A {\em quantum public-key cryptosystem\/} is a quadruple $(G,M,E,D)$, where
each component is defined as follows.
\begin{enumerate}\itemsep=0pt\parsep=0pt
\item A {\em key generator} $G$ is a (probabilistic) efficient quantum algorithm,
on input $1^n$, outputs a pair $(e,d)$. We call
$e$ {\em encryption key\/} and $d$ {\em decryption key}.
\item $M=\{M_n\}$ denotes a family of plaintext spaces, where each plaintext
is classical. We also assume that the description of $M_n$ can be
output by a polynomial-time quantum algorithm.
\item For each $n$, $(e,d)\in {\rm supp}(G(1^n))$, and $\alpha\in M_n$, 
the probabilistic {\em encryption quantum algorithm\/} $E$ and 
the deterministic {\em decryption quantum algorithm\/} $D$ satisfy that
$\Pr\limits_{E}[D_d(E_e(\alpha))=\alpha]=1$.
\end{enumerate}
%Sometimes, we denote $E_e(\cdot)$ and $D_d(\cdot)$ 
%instead of $E(e,\cdot)$ and $D(d,\cdot)$.
\end{definition}
{\bf Remark.} Note that $(e,d)$ may be a pure quantum state or a mixed quantum
state. Since mixed states are probabilistic mixture of pure states, we assume
that $(e,d)$ is a pure state without loss of generality. That is, $G$ 
probabilistically outputs a pure state $(e,d)$. 

\subsection*{Practical Requirements}
In public-key cryptography, for each key pair $(e,d)$, one receiver
and a general run of senders are involved. Thus, if $e$ and $d$ are
entangled then the receiver must keep the decryption key $d$
of the number of the senders. Though there is an application where
the above situation is useful, we consider that $(e,d)$ should not
be entangled for the standard usage of public-key cryptography.
Similarly, we consider the decryption key $d$ should be a classical state.
Thus, a key generation algorithm $G$ outputs $d$ firstly (and probabilistically)
and then outputs $e$ according to $d$. This means that there is a sub-procedure
$G'$ that, given $d$, outputs $e$. The existence of such $G'$ enables a situation
where there is a decryption key and a bunch of the corresponding encryption keys.
In this paper, we assume that quantum public-key cryptosystems satisfy the above
requirements.

\medskip

As well as the classical case,
we consider how ingredients to define the security notions should~be.

\subsection*{Attack Models}
As in the classical case, ciphertext only attack, chosen plaintext attack
and chosen ciphertext attack are considerable.
In each attack model, public key can be fed to the adversary.
Quantum public-keys seem to differ from classical ones because of the non-cloning
theorem. The nature of public-key cryptosystems is that anybody can freely access
public keys and may imply the existence of some machinery that generates
many public keys and is available even for the adversary.
Thus, we suppose that $G(1^n)$ outputs $(e,d)$ 
and $e^{\otimes {\rm poly}(n)}$ is fed to the adversary.
In this setting, ciphertext only attack is equivalent to chosen plaintext attack.

In the chosen ciphertext attack model, we give a way to make queries
in the quantum setting. The adversary queries a quantum superposition
of ciphertext to the oracle. As in the classical case, 
the amplitude of the target ciphertext in the superposition must be zero.

\subsection*{Computational Models of Adversaries}
Though we can take polynomial-size %(p-size, for short) 
circuit family
as a computational model of the adversary, we adapt a different one.
We rather take polynomial-size quantum circuit family with (non-uniform) quantum advice
as in \cite{aaronson04,ny04}. This is because in the above computational model
the distinguishability between two mixed states $\rho$ and $\sigma$
coincides with the distinguishability between $\rho^{\otimes {\rm poly}(n)}$ and
$\sigma^{\otimes {\rm poly}(n)}$ (in the computational sense).

\subsection*{Goals of Adversaries}
\begin{definition}\rm
A quantum public-key cryptosystem $(G,M,E,D)$ is said to be one-way if for every family 
$\{C_n,|a_n\rangle\}_{n\ge 1}$ of polynomial-size quantum circuits with quantum advice,
the following probability is negligible:
\[ %\begin{array}{l}
\Pr\limits_{G,E,X_n}\Bigl[
C_n(e^{\otimes {\rm poly}(n)},E_e(\alpha),|a_n\rangle)=\alpha\,\big|
(e,d)\leftarrow G(1^n);\alpha\leftarrow X_n
\Bigr] - \frac{1}{|M_n|}
\]
where $X_n$ is the uniform distribution over $M_n$.
\end{definition}

\medskip

Here, we consider the behavior of the encryption algorithm $E$.
You may consider that it is desirable for us that
the public-key state remains as it is after running $E$ and is re-usable.
(The above is not a requirement but an option.) In this case, we can write
the execution of $E_e(\alpha)$ as
\[E|e \rangle |\alpha \rangle |0\rangle = |e\rangle |\beta\rangle |\psi_{\alpha,e}\rangle,\]
where $|\psi_{e,\alpha}\rangle$ is a kind of garbage. 
In this case, the encryption $E$ must essentially produce a non-zero garbage 
state, otherwise $E^{\dagger}$ is obviously an inversion circuit.
Also note that the ciphertext is a mixed state for the receiver
since the sender in public-key cryptography cannot be an adversary
and $|\psi_{e,\alpha}\rangle$ is just local information.
On the other hand, if the encryption $E$ collapses the public-key state $e$
then the ciphertext may be a pure state.

\medskip

From now on, we define several security notions.
\begin{definition}\rm
A quantum public-key cryptosystem $(G,M,E,D)$ has {\em indistinguishability\/} if for every family
$\{C_n,|a_n\rangle\}_{n\ge 1}$ of polynomial-size quantum circuits 
with quantum advice, every polynomial $p(\cdot)$, every sufficiently large $n$, 
every distinct pair of $x,y\in M_n$, the following quantity is less than
$1/p(n)$:
\[ \begin{array}{l}\displaystyle
\Bigr| \Pr\limits_{G,E}\bigl[C_n(e^{\otimes{\rm poly}(n)},E_e(x),|a_n\rangle)=1\,
\big|
(e,d)\leftarrow G(1^n)
\bigr] \\
\hspace*{8mm}
\displaystyle
-\Pr\limits_{G,E}
\bigl[C_n(e^{\otimes{\rm poly}(n)},E_e(y),|a_n\rangle)=1\,\big|
\displaystyle
(e,d)\leftarrow G(1^n)\bigr] \Bigl| 
\end{array}
\]
\end{definition}

The above definition is just a quantum counterpart of the classical definition.
On the other hand, some care must be taken when we give a quantum counterpart
of ``semantic security''. We need to confront how to give a semantics
for the strong secrecy in the quantum computational model. While, in our setting,
information we would like to transmit is classical, we have to assume that
leakage information from the ciphertext should be either classical or quantum.
Anyway, we give two possible definitions for semantic security.

\begin{definition}\rm
A quantum public-key cryptosystem $(G,M,E,D)$ is said to be {\em semantically c-secure\/} if
there exists a (probabilistic) polynomial-time computable 
uniform quantum transformation $T$ such that
for every family $\{C_n,|a_n\rangle\}_{n\ge 1}$ of polynomial-size quantum 
circuits with quantum advice, every probability ensemble $\{X_n\}_{n\ge 1}$ 
each on $M_n$,
every polynomially bounded function $f:\{0,1\}^*\rightarrow \{0,1\}^*$,
every polynomially bounded quantum function $h:\{0,1\}^*\rightarrow {\cal H}^*$,
the following quantity is negligible with respect to $n$:
\[ \begin{array}{l}
\Pr\limits_{G,E,X_n}\Bigl[
C_n(e^{\otimes{\rm poly}(n)},E_e(\alpha),h(\alpha),|a_n\rangle)=f(\alpha)\,\big|
(e,d)\leftarrow G(1^n);\alpha\leftarrow X_n
\Bigr]
\\\displaystyle\hspace*{5mm}- \Pr\limits_{T,G,X_n}
\Bigl[C_n'(e^{\otimes{\rm poly}(n)},h(\alpha),|a_n'\rangle)=f(\alpha)\,\big|
(e,d)\leftarrow G(1^n);\alpha\leftarrow X_n
\Bigr] 
\end{array} \]
where $(C_n',|a_n'\rangle)=T(C_n,|a_n\rangle)$.
\end{definition}

\begin{definition}\rm
A quantum public-key cryptosystem $(G,M,E,D)$ is said to be {\em semantically q-secure\/} if
there exists a (probabilistic) polynomial-time computable 
uniform quantum transformation $T$ such that
for every family $\{C_n,|a_n\rangle\}_{n\ge 1}$ of polynomial-size quantum 
circuits with quantum advice, every distribution family $\{X_n\}_{n\ge 1}$ 
each on $M_n$, every pair of 
polynomially bounded quantum functions $f,h:\{0,1\}^*\rightarrow {\cal H}^*$,
the following quantity is negligible:
\[ \begin{array}{l}\displaystyle
\sum_{\stackrel{\scriptstyle (e,d)\in{\rm supp}(G(1^n))}{\alpha\in M_n}}
\!\!\!\!\!\!\!\!\!\!\!\!
\left|\langle C_n(e^{\otimes{\rm poly}(n)},E_e(\alpha),h(\alpha),|a_n\rangle)
| f(\alpha)\rangle \right|^2
\cdot \Pr[G(1^n)=(e,d)\land X_n=\alpha]
\\[2.5em]
\displaystyle
-\!\!\!\!\!\!\!\!\!\!\!\!
\sum_{\stackrel{\scriptstyle (e,d)\in{\rm supp}(G(1^n))}{
\stackrel{\scriptstyle \alpha\in M_n}{
(C_n', |a_n'\rangle)\in {\rm supp}(T(C_n,|a_n \rangle))}}}
\!\!\!\!\!\!\!\!\!\!\!\!\!\!\!\!\!\!\!\!\!\!
\left|\langle C_n'(e^{\otimes{\rm poly}(n)},h(\alpha),|a_n'\rangle) | 
f(\alpha)\rangle\right|^2
\cdot
\Pr[G(1^n)=(e,d)\land X_n=\alpha\land
T(C_n,|a_n \rangle)=(C_n', |a_n'\rangle)]
\end{array} \]
where $(C_n',|a_n'\rangle)=T(C_n,|a_n\rangle)$.
\end{definition}

We have the following equivalence among the three security notions above.

\begin{theorem}
Against the chosen plaintext attack, 
the indistinguishability, the semantic c-security, and
the semantic q-security for quantum public-key cryptosystems are all equivalent.
\end{theorem}

\def\LIMTWO#1#2{\limits_{\stackrel{\scriptstyle #1}{#2}}}
\def\LLL{\limits_{G}}
\def\MMM{\LIMTWO{G;x\in_{X_n}M_n}{r\in_U R_n}}
\begin{proof}
First, we show that the semantic q-security implies the indistinguishability.
Suppose that a quantum public-key cryptosystem $(G,M,E,D)$ does not have the indistinguishability; namely,
there exist a family $\{D_n,|b_n\rangle\}_{n\ge 1}$ of polynomial-size quantum
circuits with quantum advice and some polynomial $p(\cdot)$ such that,
for infinitely often $n$, some pair $x_n$ and $\tilde{x}_n$ both in $M_n$ satisfies 
the following:
\[\hspace*{-5mm}
\begin{array}{l}\displaystyle
\left|\Pr\limits_{G,E}
\bigl[D_n(e^{\otimes{\rm poly}(n)},E_{e}(x_n),|b_n\rangle)=1\bigr]
- \Pr\limits_{G,E}
\bigl[D_n(e^{\otimes{\rm poly}(n)},E_{e}(\tilde{x}_n),|b_n\rangle)=1\bigr]\right| > 
\frac{1}{p(n)}.
\end{array}
\]
Without loss of generality, for infinitely often $n$,
some pair $x_n$ and $\tilde{x}_n$ both in $M_n$ satisfies that
\[\hspace*{-5mm}
\begin{array}{l}\displaystyle
\Pr\limits_{G,E}
\bigl[D_n(e^{\otimes{\rm poly}(n)},E_{e}(x_n),|b_n\rangle)=1\bigr]
- \Pr\limits_{G,E}
\bigl[D_n(e^{\otimes{\rm poly}(n)},E_{e}(\tilde{x}_n),|b_n\rangle)=1\bigr] > 
\frac{1}{p(n)}.
\end{array}
\]
Here we let $X_n$ be the probability distribution satisfying that
$\Pr[X_n=x_n]=\Pr[X_n=\tilde{x}_n]=1/2$ and
$f$ be a function such that $f(x_n)=1$ and $f(\tilde{x}_n)=0$.
Now, we construct a family $\{C_n\}_{n\ge 1}$ of polynomial-size circuits
as follows.
For a given input $(e^{\otimes{\rm poly}(n)},E_e(x),|b_n\rangle)$,
$C_n$ computes $D_n(e^{\otimes{\rm poly}(n)},E_e(x),|b_n\rangle)$
and outputs the return value from $D_n$.
Then we estimate the value of
$|\langle C_n(e^{\otimes{\rm poly}(n)},E_e(x),|b_n\rangle)|f(x)\rangle|^2$ 
when $x$ is chosen according to $X_n$.
\begin{eqnarray}
\lefteqn{\sum_{\stackrel{\scriptsize {\rm supp}(G(1^n))}{\{x_n,\tilde{x}_n\}}}
\left|\langle C_n(e^{\otimes{\rm poly}(n)},E_e(x),|b_n\rangle)|f(x)\rangle\right|^2
\cdot\Pr[G(1^n)=(e,d)\land X_n=x]}\nonumber\\
&=&\Pr\LIMTWO{(e,d)\leftarrow G(1^n)}{x\leftarrow \{x_n,\tilde{x}_n\}}
\bigl[C_n(e^{\otimes{\rm poly}(n)},E_e(x),|b_n\rangle)=f(x)\bigr]
\nonumber\\
&=& \frac{1}{2}\cdot\Pr\LLL
\bigl[C_n(e^{\otimes{\rm poly}(n)},E_e(x_n),|b_n\rangle)=f(x_n)\bigr] 
+\frac{1}{2}\cdot
\Pr\LLL
\bigl[e^{\otimes{\rm poly}(n)},E_e(\tilde{x}_n),|b_n\rangle)
=f(\tilde{x}_n)\bigr] \nonumber \\
&=&\frac{1}{2}\Bigl(\Pr\LLL
\bigl[D_n(e^{\otimes{\rm poly}(n)},E_e(x_n),|b_n\rangle)=1\bigr]
+1
-\Pr\LLL
\bigl[D_n(e^{\otimes{\rm poly}(n)},E_e(\tilde{x}_n),|b_n\rangle)=1\bigr]\Bigr)\nonumber\\
&\ge &\frac{1}{2} + \frac{1}{2p(n)}.\nonumber
\end{eqnarray}
On the other hand, since $f(X_n)$ distributes over $\{0,1\}$ uniformly,
for any family $\{C_n',|a_n\rangle\}_{n\ge 1}$ of polynomial-size quantum 
circuits with quantum advice, the following holds:
\[%\begin{array}{l}\displaystyle
\sum_{\stackrel{\scriptsize {\rm supp}(G(1^n))}{\{x_n,\tilde{x}_n\}}}
\left|\langle C_n'(e^{\otimes{\rm poly}(n)},|a_n\rangle)|f(x)\rangle\right|^2
\cdot\Pr[G(1^n)=(e,d)\land X_n=x]\le \frac{1}{2}.
\]
This implies that $(G,M,E,D)$ is not semantically q-secure.

Secondly, we show that the indistinguishability implies the semantic q-security.
Suppose that there exist a family $\{C_n,|a_n\rangle\}_{n\ge 1}$ of
polynomial-size quantum circuits with quantum advice, a polynomial $p(\cdot)$,
and polynomially-bounded quantum functions $h,f$ such that, for infinitely often $n$,
the following holds:
\[ \begin{array}{l}\displaystyle
\sum_{{\rm supp}(G(1^n));M_n}\!\!\!\!\!\!\!\!
\left|\langle C_n(e^{\otimes{\rm poly}(n)},E_e(\alpha),h(\alpha),|a_n\rangle) | f(\alpha)\rangle \right|^2
\cdot\Pr[G(1^n)=(e,d)\land X_n = \alpha]\\[1em]
\displaystyle
-\sum_{{\rm supp}(G(1^n));M_n}\!\!\!\!\!\!\!\!
\left|\langle C_n'(e^{\otimes{\rm poly}(n)},h(\alpha),|a_n'\rangle)|f(\alpha)\rangle
\right|^2
\cdot\Pr[G(1^n)=(e,d)\land X_n = \alpha]
> \frac{1}{p(n)}.
\end{array}\]
In the above, $C_{n}'$ is a circuit that selects a message $\alpha'\in M_n$, 
feeds $(e^{\otimes{\rm poly}(n)},E_e(\alpha'),h(\alpha),|a_n\rangle)$ 
to $C_n$, and output the return value of $C_n$.
Also we let $|a_n'\rangle=|a_n\rangle$. 
Because of the descriptional uniformity of $M=\{M_n\}_{n\ge 1}$,
(the description of) $C_n'$ can be produced from $C_n$ by a uniform
polynomial-time quantum transformation.
Then,
\[ \begin{array}{l}\displaystyle
\sum_{{\rm supp}(G(1^n));M_n}\!\!\!\!\!\!\!\!
\left|\langle C_n(e^{\otimes{\rm poly}(n)},E_e(\alpha),h(\alpha),|a_n\rangle) | f(\alpha)\rangle \right|^2
\cdot\Pr[G(1^n)=(e,d)\land X_n = \alpha]\\[1em]
\displaystyle
-\sum_{{\rm supp}(G(1^n));M_n}\!\!\!\!\!\!\!\!
\left|\langle C_n(e^{\otimes{\rm poly}(n)},E_e(\alpha'),h(\alpha),|a_n\rangle)|f(\alpha)\rangle\right|^2
\cdot\Pr[G(1^n)=(e,d)\land X_n = \alpha]
> \frac{1}{p(n)}.
\end{array}\]
We take a message out of $M_n$ which maximizes the above difference
and let $x_n$ be the message.
By using $x_n$, we construct a circuit $D_n$ as follows.
For a given input $(e^{\otimes{\rm poly}(n)},E_e(\alpha),|a_n \rangle)$,
$D_n$ computes $C_n(e^{\otimes{\rm poly}(n)},E_e(\alpha),h(x_n),|a_n\rangle)$ 
and measures the state obtained by $C_n$ with respect to
$\Pi_0=|f(x_n)\rangle\langle f(x_n)|$ and
$\Pi_1=I - |f(x_n)\rangle\langle f(x_n)|$.
$D_n$ outputs 1 if the state is projected to $\Pi_0$ and 0 otherwise.
(Note that we may use a polynomial-time computable approximation of the projection 
instead of the exact one.)
Then we have
\[ %\hspace*{-2mm}\begin{array}{@{}l}
\Pr\limits_{G}
\bigl[D_n(e^{\otimes{\rm poly}(n)},E_e(x_n),|a_n\rangle)=1\bigr] 
-\Pr\limits_{G}
\bigl[D_n(e^{\otimes{\rm poly}(n)},E_e(\alpha'),|a_n\rangle)=1\bigr]
> \frac{1}{p(n)}.
\]
This implies that $(G,M,E,D)$ does not have the indistinguishability.

Now, we have the equivalence of the indistinguishability and the semantic q-security.
The notion of the semantic c-security is an intermediate notion between
the indistinguishability and the semantic q-security. Actually, the proof
of the equivalence between the indistinguishability and the semantic q-security
essentially includes a proof of the equivalence between the indistinguishability
and the semantic c-security.
\end{proof}

\noindent
{\bf Remark.} If we take polynomial-size quantum circuit family (without quantum advice)
as a computational model of the adversary, we do not know
whether the above equivalences still hold. This is because
we essentially use the power of quantum advice in our proof.

\begin{definition}\rm
A quantum public-key cryptosystem $(G,M,E,D)$ is said to be {\em non-malleable\/} if
there exists a (probabilistic) polynomial-time computable 
uniform quantum transformation $T$ such that
for every family $\{C_n,|a_n\rangle\}_{n\ge 1}$ of 
polynomial-size quantum circuits with quantum advice, 
every distribution family $\{X_n\}_{n\ge 1}$ each on $M_n$,
every polynomially bounded quantum function $h:\{0,1\}^*\rightarrow {\cal H}^*$,
every polynomial-size quantum circuit (with quantum advice) computable relation $R$,
the following quantity is negligible:
\[ \begin{array}{l}
\Pr\limits_{G,E,X_n}\Bigl[
C_n(e^{\otimes{\rm poly}(n)},E_e(\alpha),h(\alpha),|a_n\rangle)=E_e(\alpha')
\land (\alpha,\alpha')\in R\,\big|\,
(e,d)\leftarrow G(1^n),\alpha\leftarrow X_n
\Bigr]\\\displaystyle
\hspace*{5mm}- \Pr\limits_{T,G,X_n}
\Bigl[C_n'(e^{\otimes{\rm poly}(n)},h(\alpha),|a_n'\rangle)=E_e(\alpha')
\land (\alpha,\alpha')\in R\,\big|\,
(e,d)\leftarrow G(1^n),\alpha\leftarrow X_n
\Bigr] 
\end{array} \]
where $(C_n',|a_n'\rangle)=T(C_n,|a_n\rangle)$.
\end{definition}

\begin{theorem}
Against the chosen ciphertext attack, 
the indistinguishability and the non-malleability
for quantum public-key cryptosystems are equivalent.
\end{theorem}

\begin{proof}(Sketch)
The proof is almost similar to the proof of Theorem 3.1.

The proof that the non-malleability implies the indistinguishability
corresponds to the first half of the proof of Theorem 3.1.
In this part, we do not essentially use the power of the decryption oracle.
We have only to rephrase $f(x)=0$ and $f(\tilde{x})=1$ 
with $E_e(x)$ and $E_e(\tilde{x})$ respectively
and to let $R$ be the identical relation.

The proof that the indistinguishability implies non-malleability
corresponds to the second half of the proof of Theorem 3.1.
We can obtain, from the adversary, a ciphertext that violates the non-malleability
and invoke the decryption oracle to recover the corresponding plaintext.
Since the $R$ can be efficiently computable, then we can construct a
distinguisher that checks whether the relation holds or not.
\end{proof}

\section{Application}
A quantum public-key cryptosystem is proposed in \cite{KKNY05} and shown to have the
indistinguishability under the assumption that the graph automorphism (GA) problem
is computationally hard in the worst case. As a corollary, we have the following.

\begin{corollary}
The quantum public-key cryptosystem in \cite{KKNY05} is semantically q-secure (against chosen plaintext attack)
under the assumption that GA is a.e.-hard to compute 
by every family of polynomial-size quantum circuits with quantum advice.
\end{corollary}

%\bibliographystyle{tieice}
%\bibliography{myrefs}

\end{document}